\documentclass[12pt,manuscript]{elsart}
\usepackage{epsfig,harvard,amssymb}

\newcommand{\DD}{\displaystyle}
\makeatletter
\def\elsartstyle{%
        \def\normalsize{\@setfontsize\normalsize\@xiipt{14.5}}
        \def\small{\@setfontsize\small\@xipt{13.6}}
        \let\footnotesize=\small
        \def\large{\@setfontsize\large\@xivpt{18}}
        \def\Large{\@setfontsize\Large\@xviipt{22}}
        \skip\@mpfootins = 18\p@ \@plus 2\p@
        \normalsize
}
\makeatother

\def\bs{\bigskip}
\def\ni{\noindent}
\def\rel{relativistic \,}
\def\eg{{\it e.g.\,}}
\def\nrel{nonrelativistic \,}
\def\bibcode#1{(\texttt{#1})}

\def\url#1{{\ttfamily\def\/{/\discretionary{}{}{}}#1}}

\def\ea{et al. \,}
\def\ar{{Annu. Rev. Astron. Astrophys.} \,}

\def\apj{{ApJ.} \,}

\def\apjl{{ApJ. Lett.} \,}
\def\mn{{MNRAS} \,}

\def\physrep{Phys.Rep. \,}

\begin{document}

\begin{frontmatter}
\title{CMB Comptonization in Clusters: Spectral and Angular Power
from Evolving Polytropic Gas}

\author{Sharon Sadeh\thanksref{email}}
\address{School of Physics and Astronomy, Tel Aviv University, Tel
Aviv, 69978, Israel}

\thanks[email]{E-mail: shrs@post.tau.ac.il}

\and

\author{Yoel Rephaeli}
\address{School of Physics and Astronomy, Tel Aviv University, Tel
Aviv, 69978, Israel, \\and\\ Center for Astrophysics and Space
Sciences, University of California, San Diego, La Jolla,
CA\,92093-0424}

\begin{abstract}

The angular power spectrum of the Sunyaev-Zeldovich (SZ) effect is 
calculated in the $\Lambda$CDM cosmological model with the aim 
of investigating its detailed dependence on the cluster population, 
gas morphology, and gas evolution. We calculate the power spectrum for 
three different mass functions, compute it within the framework of 
isothermal and polytropic gas distributions, and explore the effect of 
gas evolution on the magnitude and shape of the power spectrum. We 
show that it is indeed possible to explain the `excess' power measured 
by the CBI experiment on small angular scales as originating from the 
SZ effect without (arbitrary) rescaling the value of $\sigma_8$, the 
mass variance parameter. The need for a self-consistent choice of the 
basic parameters characterizing the cluster population is emphasized. 
In particular, we stress the need for a {\it consistent} choice of the 
value of $\sigma_8$ extracted from fitting theoretical models for the 
mass function to the observed cluster X-ray temperature function, such 
that it agrees with the mass-temperature relation used to evaluate the 
cluster Comptonization parameter. Our treatment includes the explicit 
spectral dependence of the thermal component of the effect, which we 
calculate at various frequencies. We find appreciable differences between 
the \nrel and relativistic predictions for the power spectrum even for 
this superposed contribution from clusters at the full range of gas 
temperatures.

\end{abstract}

\end{frontmatter}

\section{Introduction}

The Sunyaev-Zeldovich (SZ) effect -- the change in the cosmic 
microwave background (CMB) intensity that results from Compton
scattering in a cluster of galaxies -- is a unique cosmological 
probe which is essentially independent of the cluster redshift.
The effect has already been imaged in some $\sim 60$ clusters, 
yielding important information on such quantities as the gas mass 
fraction in clusters, and the Hubble constant, $H_0$. Multi-frequency 
measurements of the effect in many more clusters are expected 
in the near future when new SZ projects -- both ground-based and 
stratospheric -- will be operational. In addition to the 
detection of the effect in a large number of clusters, planned 
sky surveys, particularly by the Planck satellite, will likely map 
the CMB anisotropy induced by the effect. The much larger SZ 
database and the measurement of this anisotropy will greatly expand 
the scope of the effect as a cosmological probe due to its strong 
dependence on the cosmological and large scale parameters, and the 
morphology and evolution of clusters. 

Extensive work in the last two decades (since it was first modeled 
[Rephaeli 1981] in the context of a simple model for IC gas evolution) 
has established the main features of SZ anisotropy in various 
cosmological, dark matter (DM), and cluster evolution models. For 
example, the SZ power spectrum was calculated in flat cold dark 
matter (\eg, SCDM, Atrio-Barandela \& M\"{u}cket 1999), $\Lambda$CDM 
(\eg, Komatsu \& Kitayama 1999), and open low-density (OCDM, Molnar 
\& Birkinshaw 2000) models (for reviews, see Rephaeli 1995a, 
Birkinshaw 1999, and Carlstrom et al. 2002). However, results for the 
power spectrum of this anisotropy and cluster number counts differ 
considerably even when calculated for the same cosmological and DM 
models, mostly due to differences in the modeling of the cluster 
mass function and gas properties, but in some cases also due to 
inconsistent choice of some of the many free parameters. There is a 
need to clarify some of these differences as part of the basic goal 
of a quantitative and {\it realistic} treatment of the SZ-induced 
anisotropy. This is clearly 
desirable
also in order to optimize strategies for the spectral and spatial 
mapping of the anisotropy in the many surveys planned by ground-based 
and satellite experiments.

The measured values of CMB intensity differences across the sky 
naturally consist of contributions from all sources of anisotropy -- 
primary, secondary, and tertiary. Most investigated is, of course, the 
primary anisotropy, originating in the very early universe, for which 
the SZ contribution could be regarded as a foreground `contamination'. 
The latter, however, is very important for better understanding of the
post-recombination universe, and -- in particular -- to the global
distribution of galaxy clusters and their physical properties. The
primary anisotropy has to be precisely known before the SZ (and other 
contributions) can be determined on angular scales of a few
arcminutes. This is particularly important now that an 
`excess' power was measured by the CBI experiment (Mason \ea 2002), at 
an estimated $3.1\sigma$ level, over what is predicted in standard models 
for the primary anisotropy at multipoles $\ell = 2000$--$3500$. This 
`excess' can possibly be due to the integrated SZ effect (Bond \ea 2002). 
The power spectrum was recently determined also by measurements with 
ACBAR (Goldstein \ea 2002) at $\sim 150$ GHz. Measurements at this and higher 
frequencies -- well beyond the CBI 26--36 GHz band -- are essential in 
order to exploit the unique spectral signature of the SZ effect as a 
powerful diagnostic tool by which it can be separated out from the 
primary anisotropy (\eg, Rephaeli 1995a, Cooray et al. 2000).

The calculation of the SZ power spectrum necessitates detailed 
modeling of the cluster mass function, intracluster (IC) gas 
properties and evolution. Consequently, the values of many parameters 
have to be specified in order to fully characterize clusters and 
their gaseous contents. Some of these parameters may be obtained from 
theoretical considerations (e.g. the critical overdensity for 
spherical collapse), while others are extracted from observations 
(\eg, cluster X-ray temperatures), or by comparing theoretical models 
with observational data, yielding important normalization factors, 
such as $\sigma_8$. This important parameter is the rms density field 
smoothed over a scale $8h^{-1}$ Mpc, where $h$ is the Hubble constant 
in units of 100 km/(s Mpc). In particular, the cluster mass-temperature 
relation and $\sigma_8$, which is usually estimated by means of the 
cluster X-ray temperature function, are coupled quantities which have 
to be consistently selected. 

Since the angular power spectrum of the SZ effect has already been 
calculated in numerous papers, we focus here on various aspects of the 
calculation which have not yet been sufficiently explored. These 
include explicit form for the temporal of IC gas, and a more general 
polytropic spatial distribution of its temperature. The description 
of the cluster population is also extended to include two mass 
functions incorporating non-spherical collapse, in addition to 
the Press \& Schechter mass function. We also discuss the need for a 
consistent choice of normalization of the cluster density and 
mass-temperature relation. In our claculation we include the full 
spectral dependence of the thermal effect which we calculate at five 
frequencies (using the exact \rel treatment) below, near, and above 
the crossover frequency (where the thermal component vanishes). These 
calculations underline the importance of using the exact relativistic 
expression for the change of intensity due to the SZ effect 
(Rephaeli 1995b), especially so near the crossover frequency. 

Our basic computational approach is essentially similar to that 
presented in a few other papers (\eg, Colafrancesco \ea 1997, Molnar 
\& Birkinshaw 2000); thus, our discussion of most aspects of the 
methodology will be brief. Rather, we will concentrate on issues that 
have attracted little or no attention. The paper is arranged as 
follows: in \S\, 2 we lay out the computational method, specifying 
the different parameters and scalings required in the calculation of 
SZ anisotropy. Results for the angular power spectrum are presented 
in \S\, 3, and discussed in \S\,4.

\section{Methodology}

The calculation of CMB anisotropy induced by the SZ effect requires 
modeling of IC gas, the cluster population, and their evolution, as 
well as the parameters of the background cosmological model. We begin 
with a brief description of the effect itself: The thermal component 
of the effect is described by the change of the spectral intensity of 
the CMB across the cluster, $\Delta I_t$. In the \nrel approximation, 
$\Delta I_t = i_{o} y g_{0}(x)$, where $i_{o}=2(kT)^3 /(hc)^2$, $T$ is 
the CMB temperature, $y=\int(kT_e/mc^2) n \sigma_T dl$ is the 
Comptonization parameter, and $g_{0}(x)$ is the well known (Sunyaev 
\& Zeldovich 1972) spectral function. Here $n$ and $T_{e}$ are the 
electron number density and temperature, $\sigma_{T}$ 
is the Thomson scattering cross section, $d\ell$ is a line element 
along a line of sight (los) to the cluster, and $x=h\nu/kT$ is the 
non-dimensional frequency. This \nrel expression for $\Delta I_t$ is 
not sufficiently accurate at high frequencies and at gas temperatures 
higher than a few keV; a more exact \rel description is then required 
(Rephaeli 1995b). As shown in the latter paper, the more exact 
\rel calculation $\Delta I_t$ differs appreciably from the 
non-relativistic expression, particularly so near the crossover 
frequency, $x_0$. This has practical ramifications since $x_0$ is 
the optimal frequency for measurement of the cluster velocity (in the 
CMB frame) along the los, and the fact that this frequency is 
extremely suitable for CMB observations, as it allows the subtraction 
of (the dominant) thermal component from the full CMB signal.

\subsection{IC Gas Properties} \label{sec:icprop} 

\ni
In all calculations it is assumed that IC gas is in a state of hydrostatic 
equilibrium in the gravitational potential of the cluster. The density 
profile adopted here is the commonly used isothermal $\beta$ model
\begin{equation}
n(r)=n_{0}\left[1+\left(\frac{r}{r_{c}}\right)^{2}\right]^{-3\beta/2}, 
\label{eq:king}
\end{equation} 
where $r_{c}$ is the core radius. For simplicity we take the cluster radial 
extent -- roughly, the virialization radius -- to be $pr_{c}=10r_{c}$. The 
core radius is scaled such that $r_{c}= 0.15\,h^{-1} \,Mpc$ for a local 
cluster with mass $10^{15}\,h^{-1}\,M_{\odot}$.

The gas mass is assumed to constitute a fraction $f$ of the total 
cluster mass. For local clusters of mass $10^{15}\,M_{\odot}\,h^{-1}$, 
we will scale to the 
nominal value $f=0.1$, in accord with recent SZ results (Carlstrom et 
al. 2002). The fact that the gas is appreciably metal-enriched, with 
typical iron abundance of $\sim 0.2-0.3$ solar, is indicative of 
galactic and intergalactic origins and varying relative 
contributions. Temporal evolution of the gas must then be taken into 
account; we do so by the parameterization 
$f(M,z)\propto M^{\eta}\,t^{\xi}(z)$, with $t$ denoting the age of 
the universe at redshift $z$. Some evidence for the form of this 
evolution was seen in cluster measurements made during the 
{\it Einstein} Medium Sensitivity Survey (EMSS; Gioia \ea 1990, 
Henry \ea 1992). Analysis of the EMSS results led Colafrancesco \& 
Vittorio (1994) to the deduction of the approximate values 
$\xi = 1.45$, and $\eta=0.2$, which we adopt here. Clearly, an 
increase of IC gas fraction with mass, and a decrease with $z$ is 
indicated by these values.

\ni 
We consider general polytropic models $T_{e} \propto n^{\gamma -1}$ 
which include the special case of isothermal distribution ($\gamma = 1$). 
The integrated $y$ parameter along a los forming an angle $\theta$ with a 
los pointing at the cluster center is given by
\begin{eqnarray}
\begin{array}{rl}
y(\theta)=\DD\frac{y_{0}}{r_{c}(M,z)}\int_{0}^{r_{1}}\DD\frac{d\ell}{\left(1+
\ell^{2}+\left(\frac{\theta}{\theta_{c}}\right)^2\right)^{\delta}} \\
=y_{0}\left[1+\left(\frac{\theta}{\theta_{c}}\right)^{2}\right]^{-\delta}r_{1}
\,  _{2}F_{1}\left[\frac{1}{2},\delta;\frac{3}{2};-\DD\frac{r_{1}^{2}}{1+
(\theta/\theta_{c})^{2}}\right].
\end{array}
\label{eq:polyt}
\end{eqnarray}
Here $y_{0}=
\DD\frac{2k_{B}\sigma_{T}}{m_{e}c^{2}}n_{0}(M,z)r_{c}(M,z)T_{0}(M,z)$,
$r_{1}=\sqrt{p^{2}-(\theta/\theta_{c})^{2}}$, $\delta = 3\beta\gamma/2$,
and $_{2}F_{1}$ is the Hypergeometric function. In the simpler
isothermal case we adopt $\beta=2/3$, and obtain
\begin{equation}
y(\theta)=\frac{y_0}{\sqrt{(1+\left(\theta/\theta_{c}\right)^{2}}}\tan^{-1}
{\sqrt{\frac{p^{2}-\left(\theta/\theta_{c}\right)^{2}}
{1+\left(\theta/\theta_{c}\right)^{2}}}}.
\label{eq:isot}
\end{equation}

\subsection{Cluster SZ Profile}

The calculation of the SZ angular power spectrum involves a number of 
stages. Here we only outline the method; further details can be found 
in previous works (\eg, Colafrancesco \ea 1997, Molnar \& Birkinshaw 
2000). The relative temperature change along a los to the cluster is
\begin{equation}
\DD\frac{\Delta T}{T}=
\left[x\coth\left(\frac{x}{2}\right)-4\right]\cdot\,y(\theta)
= s(x)\cdot \,y(\theta),
\end{equation}
where $x = h\nu/kT$ is the non-dimensional frequency, and 
$y(\theta)$ is the Comptonization parameter along a los displaced an 
angle $\theta$ from the center of the cluster. For isothermal gas 
with a King profile we have 
\begin{eqnarray}
&&y(\theta)=2\DD\frac{k\sigma_{T}}{m_{e}c^{2}}\,n(0)(z,M)\,T(0)(z,M)\,r_{c}(z,M)
\frac{1}{\sqrt{1+(\theta/\theta_{c})^{2}}} \nonumber \\
&&\cdot\tan^{-1}
{\left[p\sqrt{\frac{1-(\theta/p\theta_{c})^{2}}{1+(\theta/\theta_{c})^{2}}}
\right]},
\label{eq:ytheta}
\end{eqnarray} 
where $\theta_{c}$ denotes the angular diameter subtended by the 
cluster core. Expressions for the temperature, density and cluster 
core radius are given in Colafrancesco et al. (1994) for flat 
cosmological models, equations (6), (9) and (10), and the equivalent 
relations in equations (6) and (7) in Colafrancesco et al. (1997) for 
low-density models. For polytropic gas the corresponding expression is 
\begin{eqnarray}
\begin{array}{rl}
y(\theta)=y_{0}\left[1+\left(\DD\frac{\theta}{\theta_{c}}\right)^{2}\right]^
{-\delta}r_{1}\,_{2}F_{1}\left[\frac{1}{2},\delta;\frac{3}{2};-\DD\frac{r_{1}^
{2}} {1+(\theta/\theta_{c})^{2}}\right].
\end{array}
\end{eqnarray}
The spectral dependence of (the thermal component of) $\Delta T$,
which is given by the function $s(x)=x\coth\left(x/2\right)-4$, is
only valid at low cluster temperatures. As noted, a relativistically 
correct expression has to be used for most of the cluster temperature 
range and at high frequencies. Approximate analytic expressions for
the relativistic case have been obtained, applicable at temperatures 
$k_B\,T\leq 15\,keV$ (\eg, Itoh et al. 1998, Shimon \& Rephaeli
2003). The computation of the SZ angular power spectrum involves an 
integration over the mass-redshift distribution of the cluster
population, and thus clusters in a wide range of temperatures -- 
determined by the mass-temperature (and redshift) scaling --
contribute to the power spectrum. The predicted range of gas 
temperatures can be estimated by using the usual scaling 
$T=7.76\,\beta^{-1} 
\left(\DD\frac{M}{10^{15}\,h^{-1}\,M_{\odot}}\right)^{2/3}\,(1+z)\, 
keV$, in a flat $\Omega_0=1$ CDM model (Molnar \& Birkinshaw 2000). 
Since $\beta$ assumes values in the range $\sim 0.5 - 1$, and 
$M$ can be as high as $10^{16}\,h^{-1}\,M_{\odot}$, it follows that 
this relation yields high temperatures ($>30$ keV) at $z>1$. While a 
naive use of this scaling relation at $z>1$ is likely to be 
improper -- due (among other considerations) to fast radiative 
cooling in the high density cores of very massive clusters, if these 
existed at all already at such early times (contrary to expectations 
of hierarchical growth by merging of sub-clusters) -- it is clear 
that we have to use the relativistically correct spectral 
distribution of the SZ effect. It is nonetheless obvious that the 
mass function yields relatively low number of clusters with high 
temperatures, and that their {\it overall} contribution to the power 
spectrum is likely to be small. 

\subsection{Cluster Mass Functions}

We adopt the PS mass function (Press \& Schechter 1974) as the 
`standard' model for characterizing the cluster population. However 
reasonable and successful this function is, it still suffers from a 
number of drawbacks. From a theoretical point of view, the function 
has been derived using the oversimplified assumption of spherical 
infall, which lies on weak theoretical grounds. Its failure to deal 
with secondary infall is yet another well-known shortcoming. It has 
been shown that other theoretically derived mass functions are more 
consistent with results of N-body simulations. Specifically, Lee \& 
Shandarin (1999), and Sheth \& Tormen (1999) (hereafter LS \& ST, 
respectively) have derived mass functions incorporating non-spherical 
collapse. We calculate the SZ angular power spectrum based on all 
these three mass functions and assess the differences between the 
respective values of $C_{\ell}$.

For the mass function to correctly quantify the cluster abundance in 
the mass-redshift space, the normalization of the power spectrum of the 
density fluctuation field must be determined. This is usually specified 
in terms of $\sigma_8$, the rms density field smoothed over a scale 
$8h^{-1}$ Mpc, a free parameter in the analytic mass function expression 
whose best-fit value can be extracted from the observed X-ray temperature 
function. The temperature function describes the abundance of clusters 
$N(T)$ with temperature $T$ within a small interval $dT$; this is the 
differential form of the function. The integral form $N(>T)$ specifies 
the population of clusters with temperatures equal to or higher than 
$T$. Fitting a mass function to the temperature function requires a 
mass-temperature calibration, ascribing a temperature $T(M)$ to a 
cluster with mass $M$. The parameter $\sigma_8$ obviously depends on 
the adopted calibration. We first explain this dependence intuitively, 
then point at likely consequences of using a normalization which stems 
from an inconsistent calibration. 

Consider two temperature-mass relations (TMR) denoted by $T_1(M)$ and 
$T_2(M)$, satisfying $T_1(M)>T_2(M)$ for all $M$. Since any TMR 
predicts higher temperatures with increasing mass, i.e., the 
temperature is a monotonically increasing function of mass, the 
inverse mass-temperature relations (MTR), $M_1(T)$ and $M_2(T)$, 
must satisfy $M_1(T)<M_2(T)$ for all $T$. Consider now an observed 
temperature function $N(T)dT$, which has to be related to the mass 
function (MF). The latter is characterized by the normalization 
factor $\sigma_8$, and a fit of the presumed MF to the observed 
temperature function using a TMR may be performed with the 
normalization $\sigma_8$ as the fit parameter. The temperature 
function may be written in terms of the MF as follows:
\begin{equation}
N(T)\,dT=N[M(T)]\,\frac{dT}{dM}\,dM,
\end{equation}
and
\begin{equation}
N[M(T)]\,dM=N(T)\,\frac{dM}{dT}\,dT=N(T)\,dT\,\frac{3M}{2T},
\end{equation}
where the last equality follows from the relation $T\propto M^{2/3}$.
Given two MTRs the corresponding MF predictions are
\begin{eqnarray}
\left\{
\begin{array}{ll}
N[M_1(T)]\,dM=N(T)\,dT\,\DD\frac{3M_1(T)}{2T} \\
\\
N[M_2(T)]\,dM=N(T)\,dT\,\DD\frac{3M_2(T)}{2T}.
\end{array}
\right.
\end{eqnarray}
However, to be able to compare their predictions, the arguments of 
these functions must be identical. Defining $M_1(T)=M_2(T^{*})$ we 
obtain the following set of equations:
\begin{eqnarray}
\left\{
\begin{array}{ll}
n[M_1(T)]\,dM=n(T)\,dT\,\DD\frac{3M_1(T)}{2T} \\
\\
n[M_2(T^{*})]\,dM=n(T^{*})\,dT\,\DD\frac{3M_2(T^{*})}{2T^{*}}.
\end{array}
\right.
\label{eq:pla}
\end{eqnarray}
Now the arguments of the two MFs are identical but the MFs predictions 
themselves are different, and since $M_1(T)<M_2(T)\, \forall T$, it 
turns out that in order for the condition $M_1(T)=M_2(T^{*})$ to be 
satisfied, one must have $T^{*}<T$. This means that the lower 
expression in equation set~(\ref{eq:pla}) predicts higher cluster 
abundances. We may thus conclude that a MF derived using a MTR which 
associates a lower mass with a given temperature would predict a 
higher cluster abundance at the corresponding mass. 

The MF is to a good approximation a monotonically increasing 
function of $\sigma_8$, technically due to the fact that $\sigma_8$ 
appears in the denominator of the exponential part of $N(M)$. Only in 
the small mass range does the multiplicative term $1/\sigma_8\,$ 
dominates its behavior of the MF; in this range the MF is a 
decreasing function of $\sigma_8$ [$N(M)\propto\sigma_8^{-1}
exp(-\delta_c^{2}/2\sigma^{2}\sigma_8^{2}D_z^{2})$]. Specifically, 
$N(M)$ is an increasing function of $\sigma_8$ for $\sigma_M\cdot 
D_z<\,\delta_c$, where $\sigma_M$, $D_z$ and $\delta_c$ are the 
(unnormalized) mass variance smoothed over mass scale $M$, the linear 
growth factor of density fluctuations, and the critical density for 
spherical collapse, respectively. We can now predict the consequences 
of employing a normalization factor $\sigma_8$ extracted using a 
specific MTR, but using a different MTR for specifying the 
temperature ascribed to a cluster of mass $M$, required for the 
evaluation of the cluster $y$-parameter $\propto\,T_0$. Note that the 
SZ angular power spectrum is proportional to the product of the MF 
evaluated at specific points in the mass-redshift space and the 
temperature (through the $y$-parameter dependence). Suppose that the 
normalization parameter $\sigma_8$ is extracted by fitting a 
theoretical MF to an observed temperature function using a TMR 
$T_1(M)$, and the gas temperature used in the evaluation of the 
cluster Comptonization parameter is derived from a TMR $T_2(M)$, 
such that $T_1(M)>T_2(M)$. As seen before, the $T_1(M)$ scaling 
predicts lower cluster abundances than would be predicted employing 
the $T_2(M)$ scaling, and the $T_2(M)$ scaling attributes lower 
temperatures to a given mass. This could lead to an appreciable 
underestimation of the SZ $C_\ell$'s. In the opposite case, when 
$T_2(M)$ is used to extract $\sigma_8$, and $T_1(M)$ to evaluate the 
temperature, $C_\ell$ will likely be overestimated. This 
inconsistency underlines the need for a unique choice of the TMR.

\subsection{Angular Power Spectrum}

The angular power spectrum of the SZ effect (\eg, Molnar \& Birkinshaw
2000) in the cluster population is characterized by 
\begin{equation}
C_{\ell}=\int_{z} r^{2}\,\frac{dr}{dz}\int_{M} N(M,z)\,
\zeta_{\ell}(M,z)\,dM\,dz,
\end{equation}
where $N(M,z)$ is the MF, and $r$ is the comoving radial distance. The 
function $\zeta(M,z)$ is the angular Fourier transform of the spatial 
profile of $\Delta T$, which in the \nrel limit is given in terms of 
the profile of the Comptonization parameter (eq. 5). In our calculations 
we consider clusters in the mass range of 
$10^{13}-10^{16}\,M_{\odot}\,h^{-1}$, but discuss how the results change 
when a narrower range is taken. The redshift integration interval -- 
determined by the earliest epoch when clusters can be assumed to have formed 
and virialized -- is uncertain and clearly depends on the cosmological model.
\bs

\section{Results and Discussion}

\ni
The `standard' model in our calculations of the SZ angular power spectrum 
is the currently favored flat, vacuum-dominated universe. The model is 
characterized by the following set of cosmological parameters: $\Omega_m=0.3, 
\Omega_{\Lambda}=0.7, \Omega_B=0.05,n=1, h=0.7$. Clusters are assumed to 
be distributed according to the PS MF, have isothermal gas over a spherical 
region with a radius of $10r_c$, with the gas mass fraction that scales with 
the cluster mass and evolves in time as $f(M,z)\propto M^{0.2}\,t(z)^{1.45}$. 
We have also used two other recently proposed MFs, and generalized the gas 
distribution to a polytrope with a range of values for the index. In the 
calculations we have also considered the case of no gas evolution. The 
various cases of gas parameters for a given cluster are treated 
self-consistently by keeping the total mass and the gas mass fraction 
fixed as the polytropic index is varied. We carried out the 
calculations both without and with the SZ spectral dependence (the latter 
only for our standard model) evaluated at five observing frequencies -- 
taken to correspond to the four MITO channels (De Petris \ea 2002) and 
the 545 GHz channel of the HFI on the Planck satellite.

\begin{figure}[t]
\centering
\epsfig{file=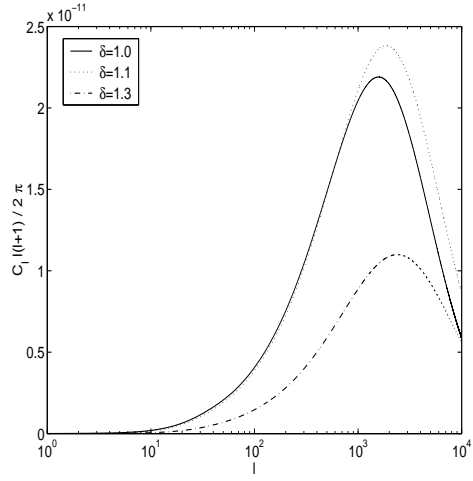, height=6.4cm, width=6.2cm, clip=}
\caption{SZ angular power spectrum for isothermal ($\delta=1.0$) gas 
and two polytropic temperature profiles with $\delta=1.1$ and $1.3$.}
\label{fig:poli}
\end{figure}

\begin{figure}
\centering
\epsfig{file=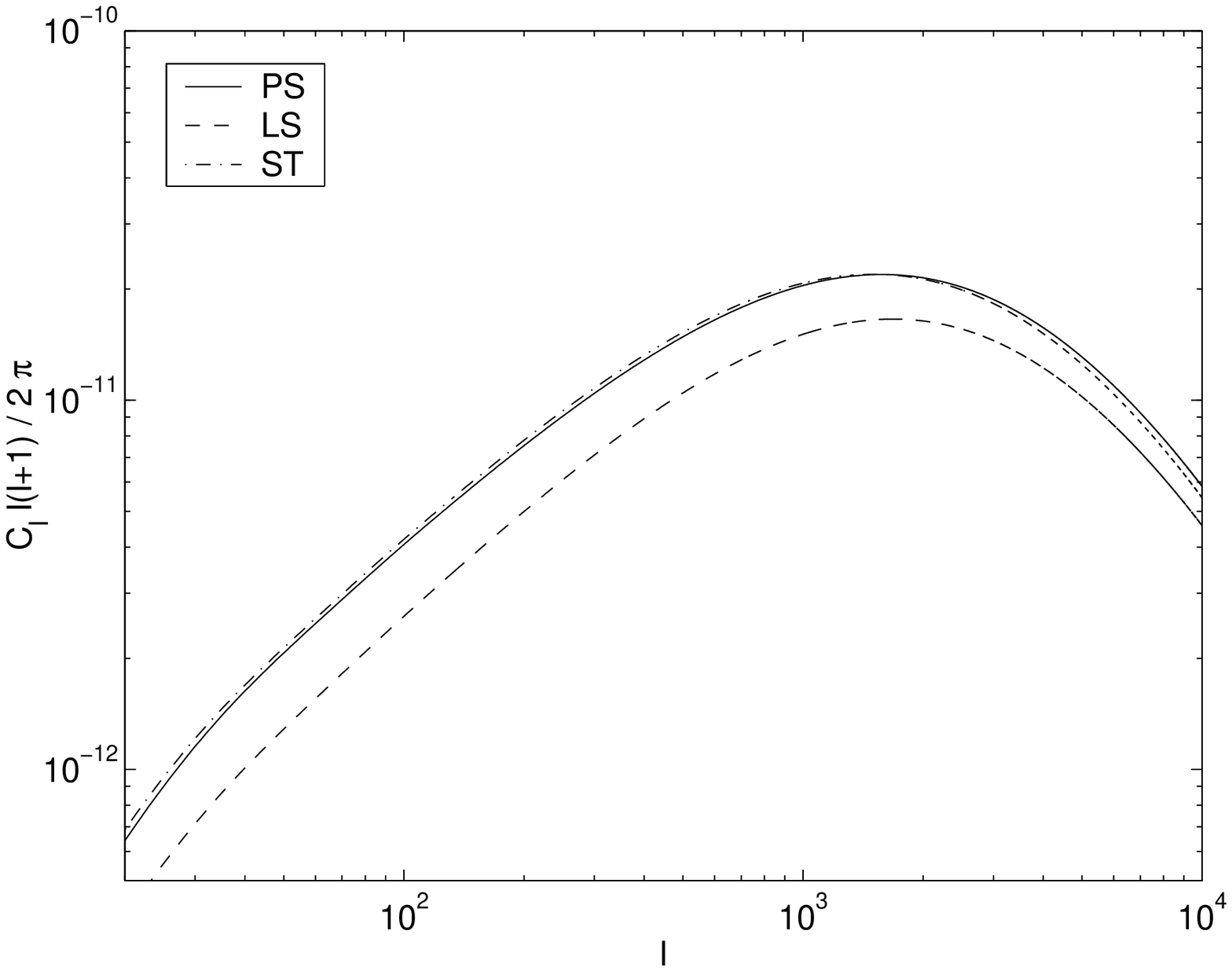, height=7.4cm, width=6.2cm, clip=}
\epsfig{file=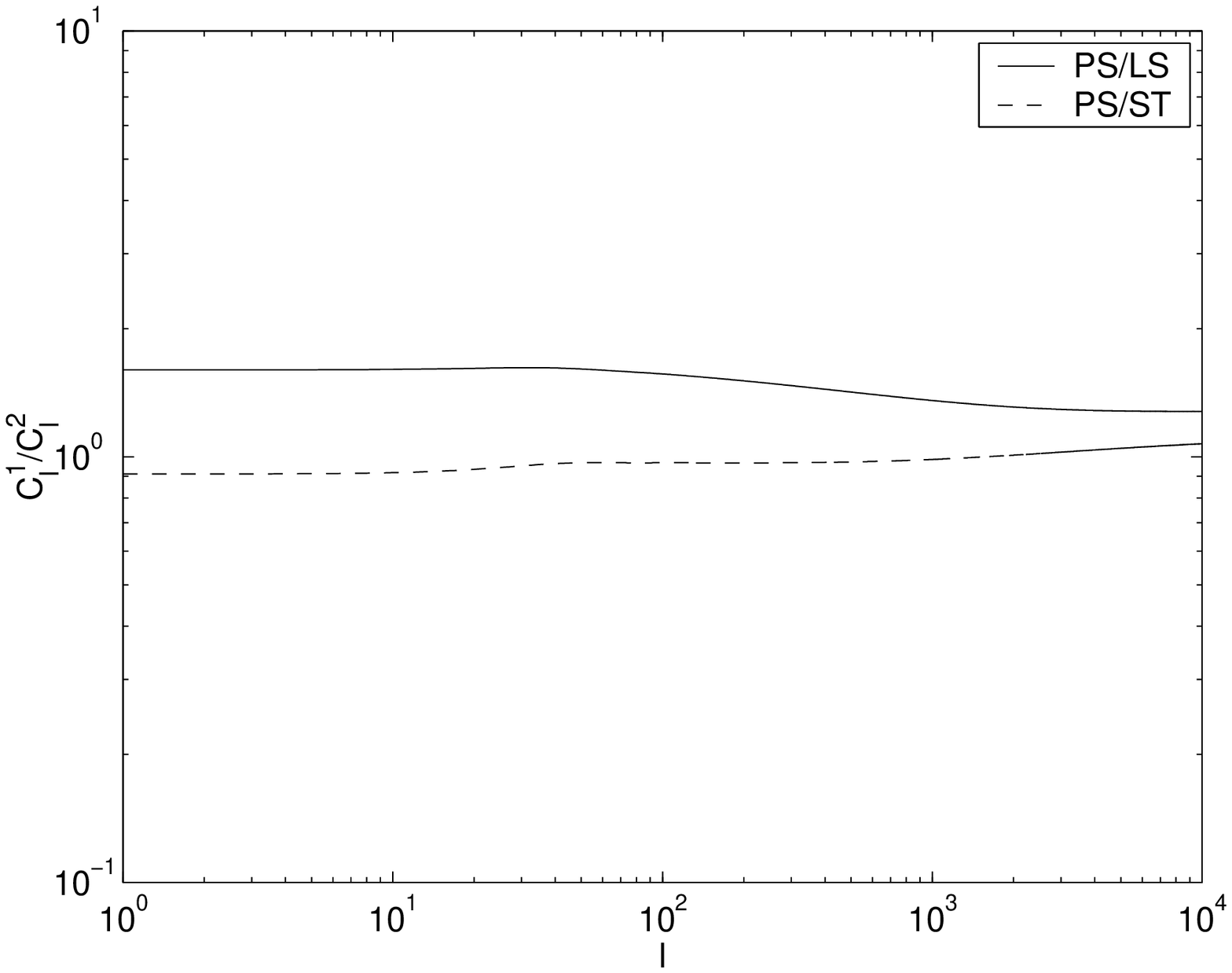, height=7.4cm, width=6.2cm, clip=}
\caption{SZ power spectrum calculated using the PS, LS and ST mass 
functions}
\label{fig:mf}
\end{figure}

\begin{figure}
\centering
\epsfig{file=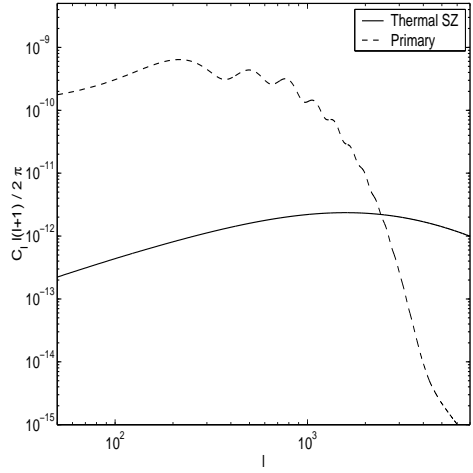, height=6.4cm, width=6.2cm, clip,}
\caption{SZ and primary anisotropy power spectra in the $\Lambda$CDM 
models}
\label{fig:szprim}
\end{figure}

As mentioned in section 2.3, a mass-temperature scaling is needed in
order to fit a MF to the observed cluster X-ray temperature function, so
as to obtain the normalization $\sigma_8$. We have adopted the mass-temperature
relation and the deduced values of $\sigma_8$ obtained by Wu (2000) from 
fits of the LS and ST MF to the observed temperature function derived 
by Viana \& Liddle (1999). The latter authors relaxed the assumption of 
spherical gravitational collapse, and derived analytic expressions within 
the framework of non-spherical collapse. (We note that in these models a 
halo may begin to form when an overdensity collapses along one of the 
principal axes of the deformation tensor. The earlier formation of clusters 
would delay the active formation epoch in comparison with the PS case 
[Wu 2000], i.e., a cluster with present virial mass $M$ has a higher 
probability of forming at a lower redshift than in the PS models.) For our 
choice of the standard model parameters, the values of $\sigma_8$ are 
$0.974, 0.839, 0.913$ for the PS, LS and ST MFs, respectively. This 
parameter was recently deduced also from the SDSS and WMAP CMB anisotropy 
databases, yielding somewhat lower values than we have 
adopted here (Melchiorri \ea 2002, Bennett \ea 2003); however, the 
range of uncertainties in the various deduced values of $\sigma_8$ is 
wide and certainly includes the above three values that we use here. 
Results for the power spectrum for the standard model with full, partial, 
and no gas evolution (to be specified below) are plotted in 
figure~\ref{fig:gasevol}. Note that here and in the following figures the 
power spectrum is expressed in its dimensionless form, namely it is 
$\Delta T/T_0$ (rather than $\Delta T$) that is expanded in Fourier space. 
In Figure~\ref{fig:poli}, power spectra are shown for two polytropic models 
with indices $\gamma=1.1, 1.3$, and for the standard isothermal model. The 
predictions from the PS, LS and ST mass functions are shown in 
figure~\ref{fig:mf}. The primary and SZ induced anisotropy power spectra 
are shown in figure~\ref{fig:szprim}.

The SZ power spectrum in the standard model and in the models with the 
LS and ST MFs are shown in figure~\ref{fig:mf}. Note that all three curves 
lie close to each other, particularly the PS and ST MFs. The LS MF yields 
lower values for the power spectrum; as can be seen in the right 
panel of figure~\ref{fig:mf}, this amounts to a reduction of the power 
spctrum by $\sim 50\%$, somewhat higher at low multipoles, and lower by 
$\sim 10\%$ at low multipoles; this curve virtually coincides with the PS 
prediction at higher multipoles. Apparently, when these MFs are fit to the 
same observational data by using the same mass-temperature relation, 
differences among the power spectra will be minor. This is in particular 
important due to the fact that the MF is very sensitive to the input 
parameters, in particular to those appearing in its exponential part. A 
substantial amount of effort has been made over recent years to improve the 
PS model by resolving some of its problematic features, and to obtaining more 
realistic MFs. It appears that at least for the task of calculating SZ power 
spectra, using these functions instead of the PS MF does not affect much the 
results, particularly so when compared with other sources of uncertainties.

In all the models considered here the SZ power spectrum exhibits the same 
general behavior: a monotonic increase from low levels on large scales 
(low $\ell$), with a steep rise to a peak centered around 
$\ell \sim 1500-3000$ (depending on the model), followed by a steep decline. 
Differences between the respective curves are manifested in the orientation 
of the curve relative to the multipole axis and the magnitude -- or height 
-- reached at the peak. This characteristic curve stems from the angular 
Fourier transform of the Comptonization parameter. For small values of 
$\ell$ the Fourier transformed profile is essentially constant in magnitude; 
the parabolic shape is due to multiplication by $\ell(\ell+1)$. On smaller 
scales the Fourier transformed profile is exponentially damped, becoming 
sensitive to the internal structure of the cluster. This gives rise to the 
steep fall of the power spectrum; the peak of the curve roughly represents 
the transition between large and small scales, or between unresolved and 
resolved clusters. Note that the power spectrum is calculated only for the 
thermal component of the SZ effect; the related anisotropy induced by the
kinematic component has been treated in various papers (\eg, Cooray
2001). The magnitude of the integrated contribution of this component 
is second order in the cluster velocity (in the CMB frame), and thus very 
much smaller than the respective contribution of the main, thermal component.

Gas evolution is a major factor determining the shape of the power spectrum, 
as is apparent from figure~\ref{fig:gasevol}; with our adopted form of gas 
evolution the curve peaks at $\ell\sim 1400$, reaching a magnitude of 
$\sim 2.2\cdot10^{-11}$. Without gas evolution the peak shifts considerably 
to $\ell\sim 8000$, where it reaches the much higher value of 
$\sim 2.4\cdot10^{-10}$. Comparison of the two curves reveals that the 
higher the multipole, the greater is the contrast between the powers of 
the non-evolving and evolving gas. This stems from the higher gas densities 
in distant clusters predicted in the non-evolutionary gas scenario, which 
gives rise to an increase of power on smaller scales, reflecting the small 
angular sizes of distant clusters. It is also interesting to calculate the 
power spectrum in two 'partial' gas-evolving models, the first chosen to 
correspond to $\xi=0$ and $\eta=0.2$, and the second to $\xi = 1.45$ and 
$\eta=0$ -- namely removing from the gas mass fraction the redshift and mass 
dependences, respectively -- in order to gauge the effect of the range of 
values of these parameters on the calculated power. For example, removing 
the redshift dependence results in a considerable dependence of the results   
on the mass lower limit, since -- as can be seen in figure~\ref{fig:gasevol} 
-- the non-evolving gas model still dominates the partially evolving gas 
model by a factor of $\sim 17$ at the highest multipoles. 
This maximal factor corresponds to the lowest cluster mass considered in 
the calculation, $M=10^{13}M_{\odot}\,h^{-1}$; this factor would have 
been smaller had we taken a higher mass cutoff.

The $\eta=0$ curve provides insight into the redshift range where most 
contributing to the anisotropy. Since the power spectrum in the model with 
non-evolving gas is only a factor of $\sim 2$ larger (on average) than that 
of evolving gas, we can evaluate the maximal redshift at which the 
contribution is still appreciable: In a flat universe we have 
\begin{equation}
\DD\frac{t}{t_{0}}=\DD\frac{\ln\left[\frac{\Omega_{\Lambda}^{1/2}+
\left[\Omega_{\Lambda}+\Omega_m(1+z)^{3}\right]^{1/2}}
{\Omega_m^{1/2}(1+z)^{3/2}}\right]}{\ln\left[\frac{\Omega_{\Lambda}^{1/2}+1}
{\Omega_m^{1/2}}\right]}.
\label{eq:tt0lamb}
\end{equation}
Now, the ratio between the power spectra excluding gas evolution and 
including only its temporal dependence is, $(t/t_{0})^{-\xi}$, and with 
the best fit value $\xi=1.45$, we determine the maximal factor this term 
contributes to the power spectrum in the non-evolving scenario to be 
$(t/t_{0})^{-1.45} \sim \sqrt{2}$. Employing equation~(\ref{eq:tt0lamb}) 
to solve for the redshift yields $z\approx 0.24$. The relatively low value 
of this redshift may seem surprising as we may expect high-redshift objects 
to contribute considerably to the power spectrum when there is no gas 
evolution. Instead, we only see a slight enhancement of its intensity, which 
peaks around a rather low value. One may then conclude that the ever 
decreasing cluster population at higher redshifts is responsible for this 
effect, and that the scaling of IC gas with mass dominates over its temporal 
change.

It is unrealistic to expect that thermal conductivity in the gas is 
sufficiently high for it to be isothermal over the full extent of the 
gas distribution. A more general distribution is that of polytropic gas 
a polytropic equation of state, $P\propto\rho^{\gamma}$, for which the 
Comptonization parameter $y$ scales as a los integral of $n^{\gamma}$. 
Clearly, for isothermal gas and $\beta=2/3$, $\delta=1$ (eq. 2). We 
consider two values of the polytropic index: $\gamma=1.1, 1.3$, 
in addition to the isothermal case. Using the hydrostatic equation, the 
central temperature is 
\begin{equation}
T_{0}=\DD\frac{G\mu
m_{H}\left(1+p^{2}\right)^{\delta}}{k_{B}\rho_{0}\left[(1+p^{2})^{\delta}-
1\right]}\DD\frac{\psi'(p,\beta)}{\omega(p,\beta)};
\end{equation}
the functions $\psi'(p,\beta)$ and $\omega(p,\beta)$ are defined in 
Colafrancesco \ea (1994). While the central temperature is not sensitive to 
the exact value of $\gamma$, the Comptonization parameter has a steeper
distribution in the polytropic case, resulting in a substantially different 
power spectrum than that of isothermal gas. Figure~\ref{fig:poli} 
describes the SZ angular power spectrum in the isothermal and the two other 
polytropic cases. 
While there is a slight increase of the power for $\gamma=1.1$ (as compared 
to the isothermal case), the peak value for $\gamma=1.3$ is a factor
$\sim 2$ lower with respect to the isothermal gas. 
The peaks shift to higher values of $\ell$ in both ploytropic cases. 
These changes result from a steeper Comptonization parameter due to a more 
compact cluster, leading to smaller cluster size and thus an increase of power 
on smaller scales. The peaks of the two curves for polytropic models 
with $\gamma > 1$ occur in the range $\ell\approx 2000-3000$; 
the steeply falling part of the profiles shifts further to the right along 
the multipole axis. The difference between the magnitudes in the two 
polytropic cases is likely due to an interplay between two competing 
effects: 
the steeper Comptonization parameter assumes higher values in the central 
cluster regions, but it assumes these values along shorter distances. 
Therefore, with polytropic indices slightly above unity, there 
is an increase of magnitude, where the steeper profile dominates over the 
somewhat shorter distances along 
lines of sight with constant values of $y$, while 
for higher values of $\gamma$ the impact of the reduced pathlength dominates.

\begin{figure}
\centering
\epsfig{file=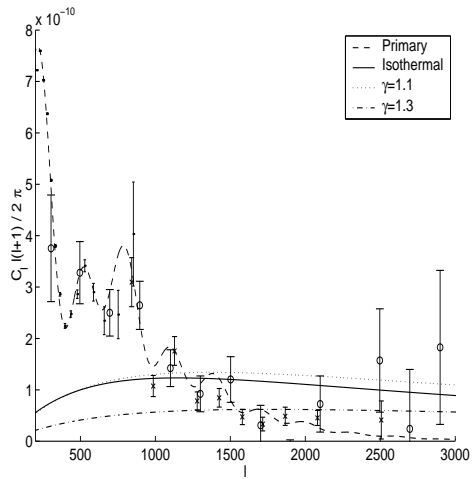, height=6.4cm, width=6.2cm, clip}
\caption{Recent results from the CBI, ACBAR, and WMAP experiments, together 
with the angular power spectrum of the primary anisotropy as calculated
by CMBFAST, and the SZ power spectrum in isothermal and polytropic
models. WMAP, ACBAR, and CBI data are marked by dots, crosses, and empty 
circles, respectively.}
\label{fig:cbisz}
\end{figure}

In figure~\ref{fig:cbisz} we plot the primary anisotropy, calculated using 
the CMBFAST code, and show the results from WMAP, CBI, and ACBAR 
measurements, as well as SZ power spectra in the isothermal and polytropic 
cases considered above. We have included the spectral dependence which we 
calculated at a frequency of $31\,GHz$, as appropriate for the CBI experiment. 
ACBAR measurements (Kuo \ea 2002) were at a frequency of $150\,GHz$; the 
data points are shown as reported (in the latter paper). Were it known that 
the signal measured at high multipoles is largely due to the SZ effect, we 
could have easily scaled the ACBAR data to the lower CBI frequency by the 
theoretically predicted spectral factor of $\sim 4.2$ (assuming a mean 
gas temperature of ~7 keV). This would have resulted in an enhancement of 
the ACBAR measured power over the data collected at the actual observing 
frequency.

The `excess' power detected by the CBI experiment at high multipoles
(Mason et al. 2002) may possibly be attributed to SZ signal. It has 
been suggested that this measured `excess' is likely to require a 
modification of $\sigma_8$, in order to increase the cluster 
population, as predicted by the PS mass function. However, such a 
modification is not that simple to make; $\sigma_8$ is usually estimated
by fitting a theoretical mass function to the observed temperature
function of X-ray clusters; therefore, its value is set by the fit. 
Moreover, different analyses yielded various values which span the rather 
broad range, $0.7\lesssim\sigma_8\lesssim 1$. An arbitrary modification of 
this parameter would have consequences on the mass-temperature relation, 
which might be in total disccordance with the original mass-temperature 
relation employed in order to extract $\sigma_8$ in the first place. Our 
results seem to be consistent with the CBI measurements, thus circumventing 
the need for a modification in this parameter. However, even 
if this was not the case, discrepancies between the two sets of results may 
be explained by other means than the normalization parameter $\sigma_8$. In
particular, the not fully known properties of IC gas provide a certain degree 
of freedom in adjusting the theoretically predicted level of the anisotropy. 
For exmaple, as we have shown above, the anisotropy depends apprecaibly on 
the evolution of IC gas; moderate (or no) evolution results in a 
considerable increase of the power spectrum at high multipoles. This is 
particularly relevant given the observational indications that there is 
weak or no evolution of the X-ray luminosity function of clusters. 

In figure~\ref{fig:szprim} we compare the SZ power spectra obtained in the 
$\Lambda$CDM model with the corresponding power spectra of the primary 
anisotropy. The latter was computed using the CMBFAST code (Zaldarriaga \& 
Seljak 1996). As can be seen from the figure, the SZ power spectrum begins 
to dominate over the primary anisotropy at multipoles of $\ell\sim 2000$, 
but even at lower multipoles ($\ell\gtrsim 1500$) the magnitude of the 
SZ power already constitutes a fraction ($\gtrsim 10\%$) of the primary 
power. Furthermore, these results were obtained with constant spectral 
function, $s(x)=1$. Its value in the Rayleigh-Jeans region is $-2$, but 
it can be much larger; for example at the highest frequency of the HFI 
instrument of Planck ($\sim 857$ GHz), $s(x) \simeq 11$, so -- from merely 
a theoretical point of view the -- there is a clear advantage in measuring 
the power at high frequencies. 

\begin{figure}
\centering
\epsfig{file=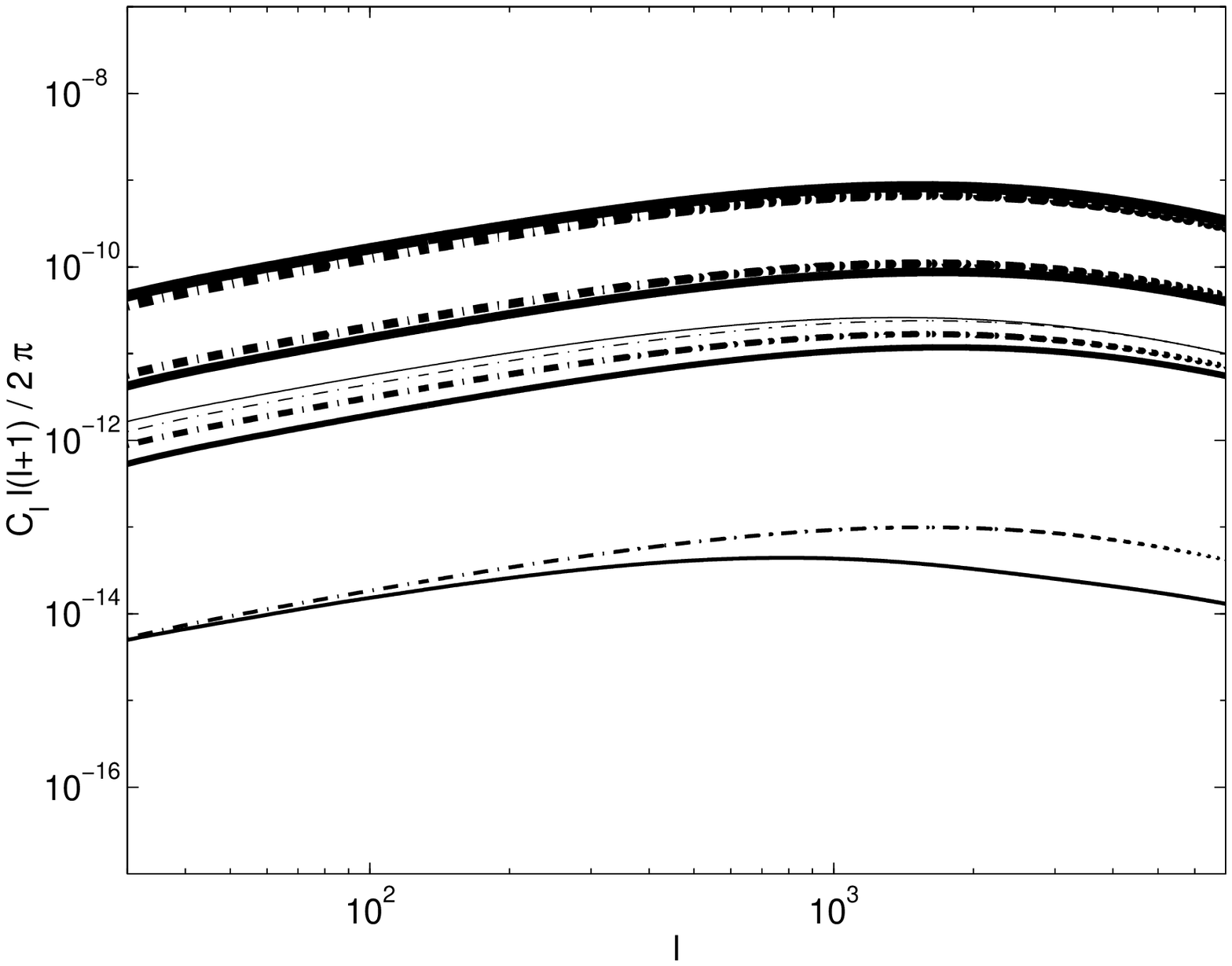, height=6.4cm, width=6.2cm, clip}
\epsfig{file=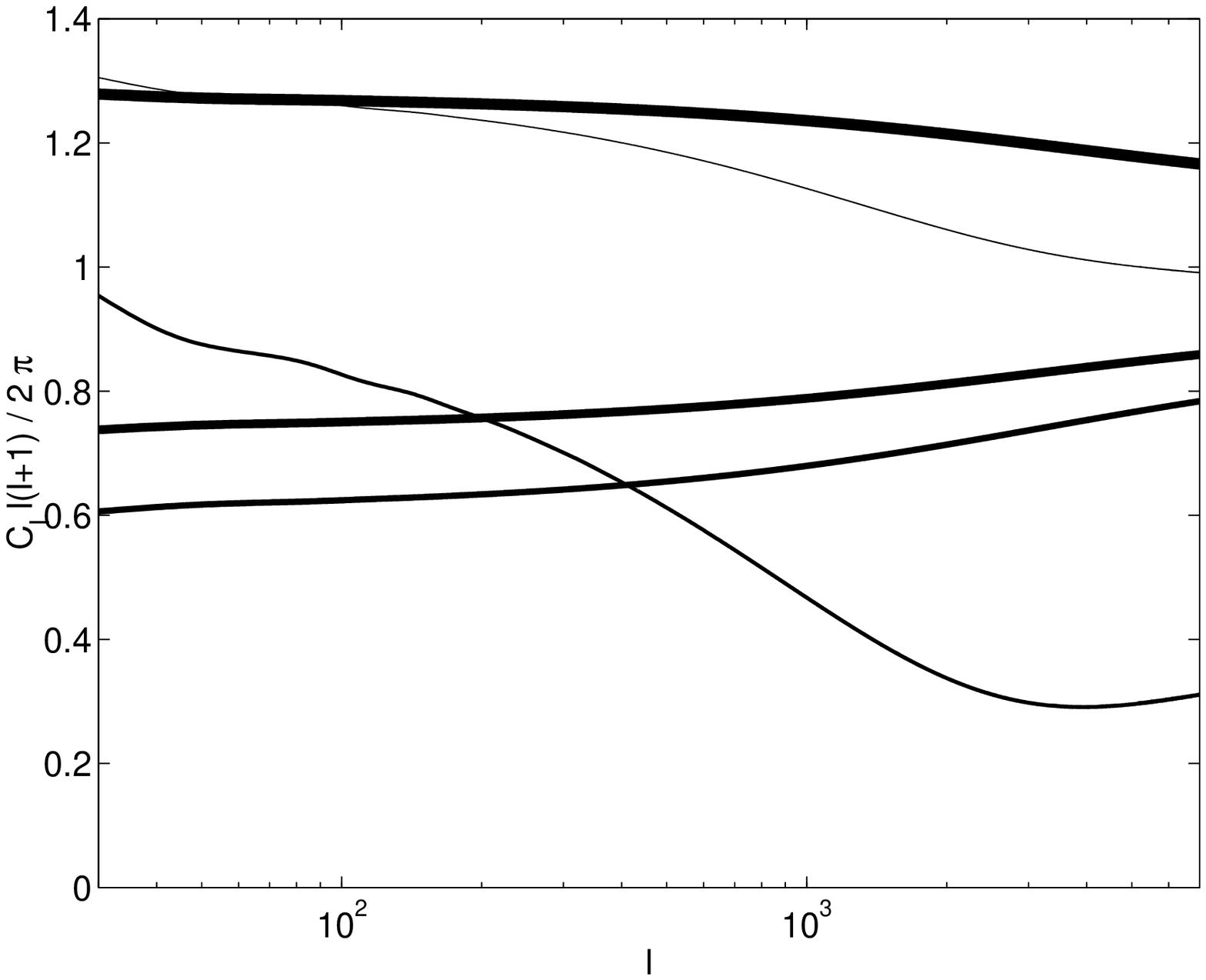, height=6.4cm, width=6.2cm, clip}
\caption{Power spectra calculated using the non-relativistic and exact 
relativistic SZ intensity change are shown in the left panel. The 
spectral dependence was calculated at five frequencies, $x=2.5, 3.9, 
4.8, 6.2, 9.6$; each pair of adjacent solid and dash-dotted lines 
correspond to the relativistic and non-relativistic calculation, 
respectively. Line thickness increases with the value of $x$. Ratios 
of relativistic to non-relativistic power at each frequency are shown 
in the right panel.}
\label{fig:rsz1}
\end{figure}

\begin{figure}
\centering
\epsfig{file=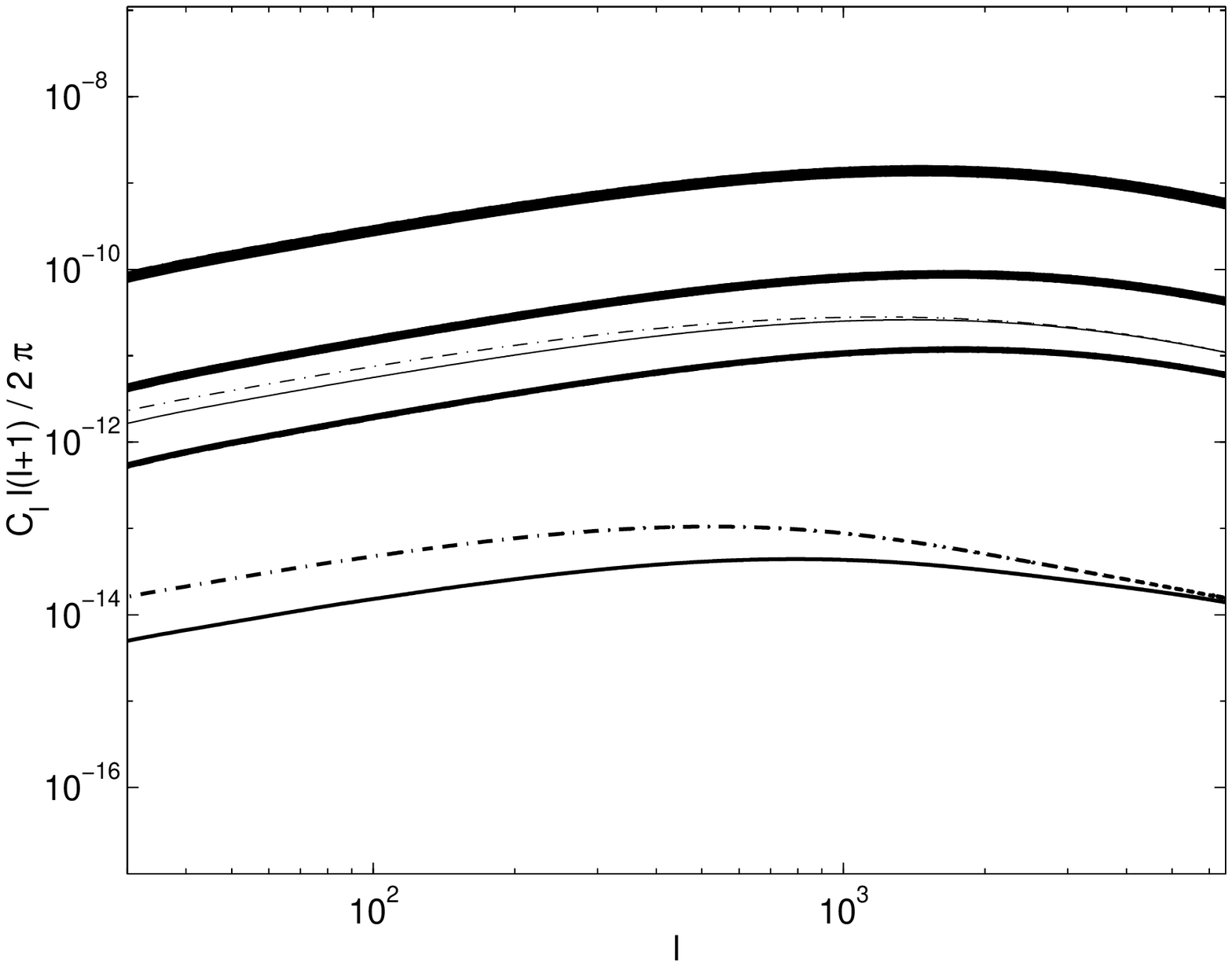, height=6.4cm, width=6.2cm, clip}
\epsfig{file=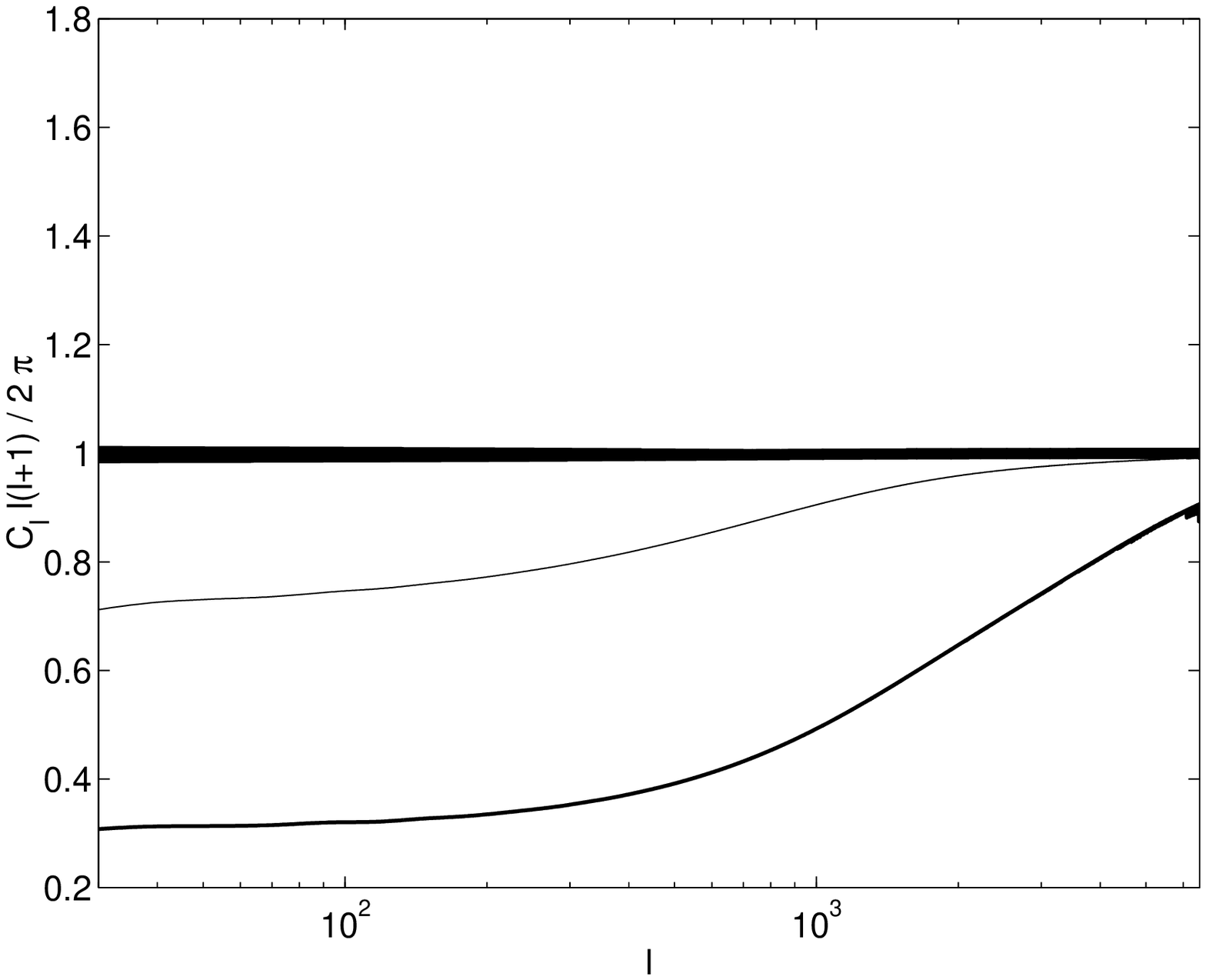, height=6.4cm, width=6.2cm, clip}
\caption{Comparison of the results from the exact \rel and approximate 
analytic expressions. Solid lines show the power spectrum calculated 
using the exact \rel calculation of Rephaeli (1995b) for the spectral 
function, whereas the dash-dotted lines show the corresponding results 
when the analytic expression of Itoh et al (1998) is used. Ratios of 
the results from the exact relativistic calculation to those from the 
analytic approximation are shown in the right panel; line thickness
corresponds to frequency as in the previous figure.} 
\label{fig:rsz2}
\end{figure}

The power spectrum was calculated at five frequencies, four of which 
correspond to the channels, $x= 2.5, 3.9, 4.8, 6.2$, and a 
high frequency -- the highest HFI channel with reasonably adequate 
sensitivity to SZ measurements aboard the PLANCK satellite, $\nu=545\,GHz, 
x=9.6$. To assess the degree of importance of using the relativistic SZ 
expression we plot the SZ angular power spectrum obtained using the
non-relativistic expression, the exact relativistic expression, and the 
approximate analytic expression of Itoh et al. (1998):
figure~\ref{fig:rsz1} compares the SZ angular power spectrum calculated  
with the non-relativistic expression of the spectral distortion with that
calculated using the exact relativistic expression. The results demonstrate 
(again) the need for using the exact \rel expression. The approximate analytic 
expression is of limited use near the the crossover frequency, as shown in 
figure~\ref{fig:rsz2}. For example, at $x=2.511$ and $x=3.907$ and high 
values of $\ell$, it yields values that are $25\%-200\%$ larger than those 
calculated using the exact \rel treatment.

\begin{figure}
\centering
\epsfig{file=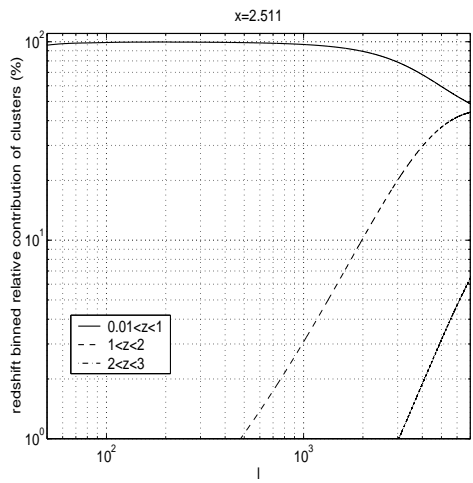, height=6.4cm, width=6.2cm , clip}
\caption{Relative contribution to the overall power spectrum, 
at $x=2.5$, by clusters at three redshift intervals.} 
\label{fig:clz}
\end{figure}

Finally, to determine the redshift distribution of the SZ power spectrum 
contribution of clusters, we divide the redshift integration range 
$(0.01\le z\le 6)$ into six intervals, and plot in figure~\ref{fig:clz} 
the fractional contribution of each interval to the overall power spectrum 
at $x=2.5$ (the corresponding plots at the other frequencies do not differ 
substantially). As can be seen in the figure, the contribution to the 
power spectrum from redshifts $z>1$ is negligible at multipoles 
$\ell\lesssim 1000$. The contribution of clusters at $1<z<2$ increases 
from $\sim 10\%$ at $\ell\sim 2000$, to as much as $40\%$ at the highest 
multipoles. Clusters at $2<z<3$ contribute only at the highest multipoles 
at a level $\leq 7\%$, and the contribution of clusters at still higher 
redshifts is $< 1\%$, which is the reason for its absence in the figure. 
These results agree with the analysis done based on the partial 
gas-evolution scenario discussed earlier in this section. 
These results also imply that even though high gas temperatures are 
predicted at $z>1$ by the virial mass-temperature relation, which is based 
on a  naive redshift scaling, the impact of this unrealistic relation on 
our results is minimal since the contribution of high temperature clusters 
is negligible.

\section{Conclusion}

One of the main goals of our work has been to understand the reasons 
for the substantial variance among previous predictions for the power 
spectrum of the SZ-induced anisotropy. Among the various works on the 
power spectrum which employ analytic forms of the mass function to 
calculate the power spectrum in the context of a given cosmological and 
DM model, the main reason for quantitative differences was shown to be 
a choice of the cluster mass-temperature relation that is inconsistent 
with value of $\sigma_8$ which is adopted in the calculation. 
We have taken explicit account of the evolution of the gas fraction in 
the context of a simple model, and explored how the power spectrum depends 
on the index of a polytropic equation of state. Our results are therefore 
self-consistent, more general in the descriptionof the properties of the 
gas than previous treatments, and include detailed predictions of the 
power spectrum at five frequencies based on a relativistic calculation 
of the spectral distribution of the SZ effect.

\newpage
\def\ref{\par\noindent\hangindent 20pt}

\newpage
\begin{figure}
\centering
\epsfig{file=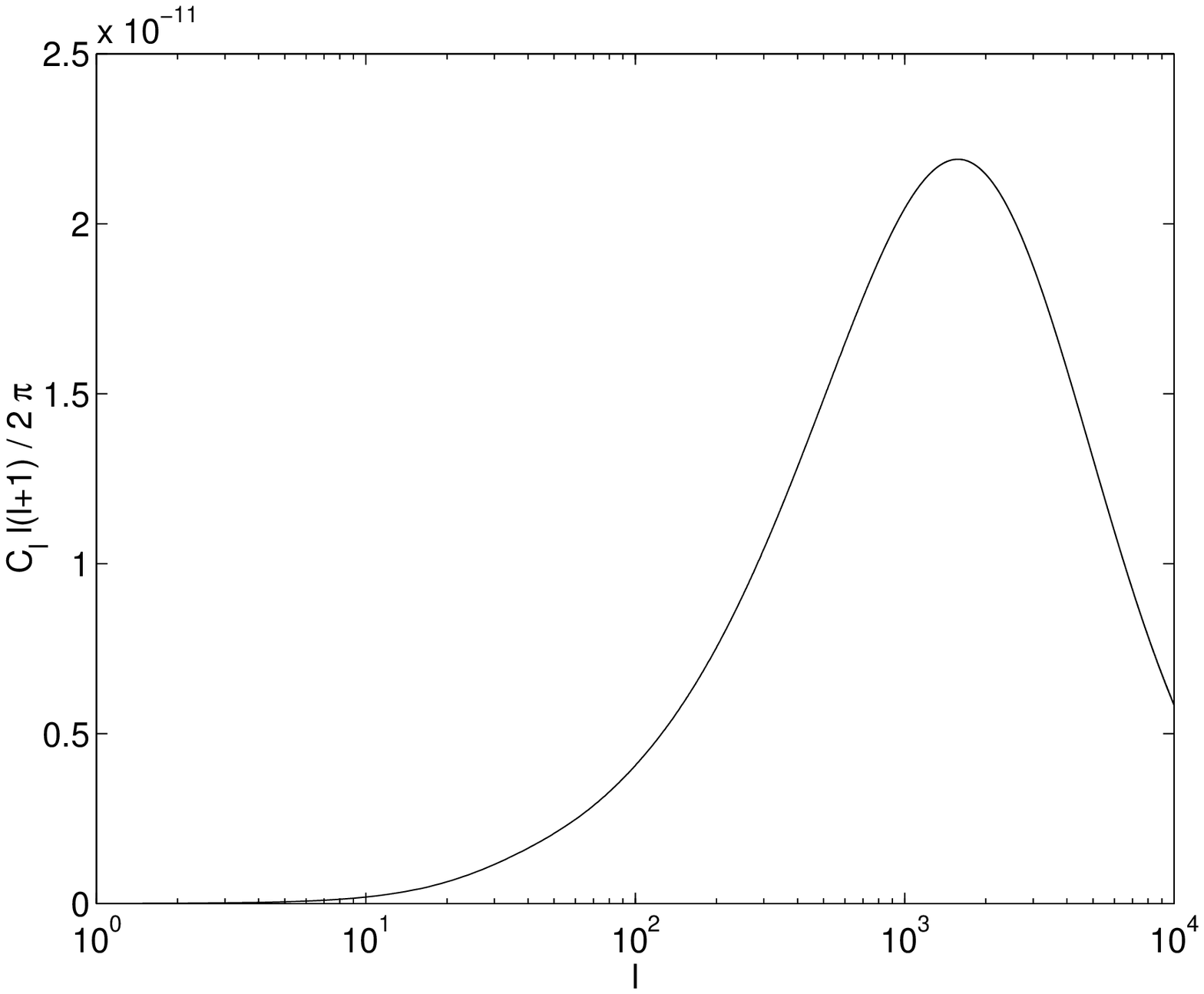, height=6.4cm, width=6.2cm, clip=}
\epsfig{file=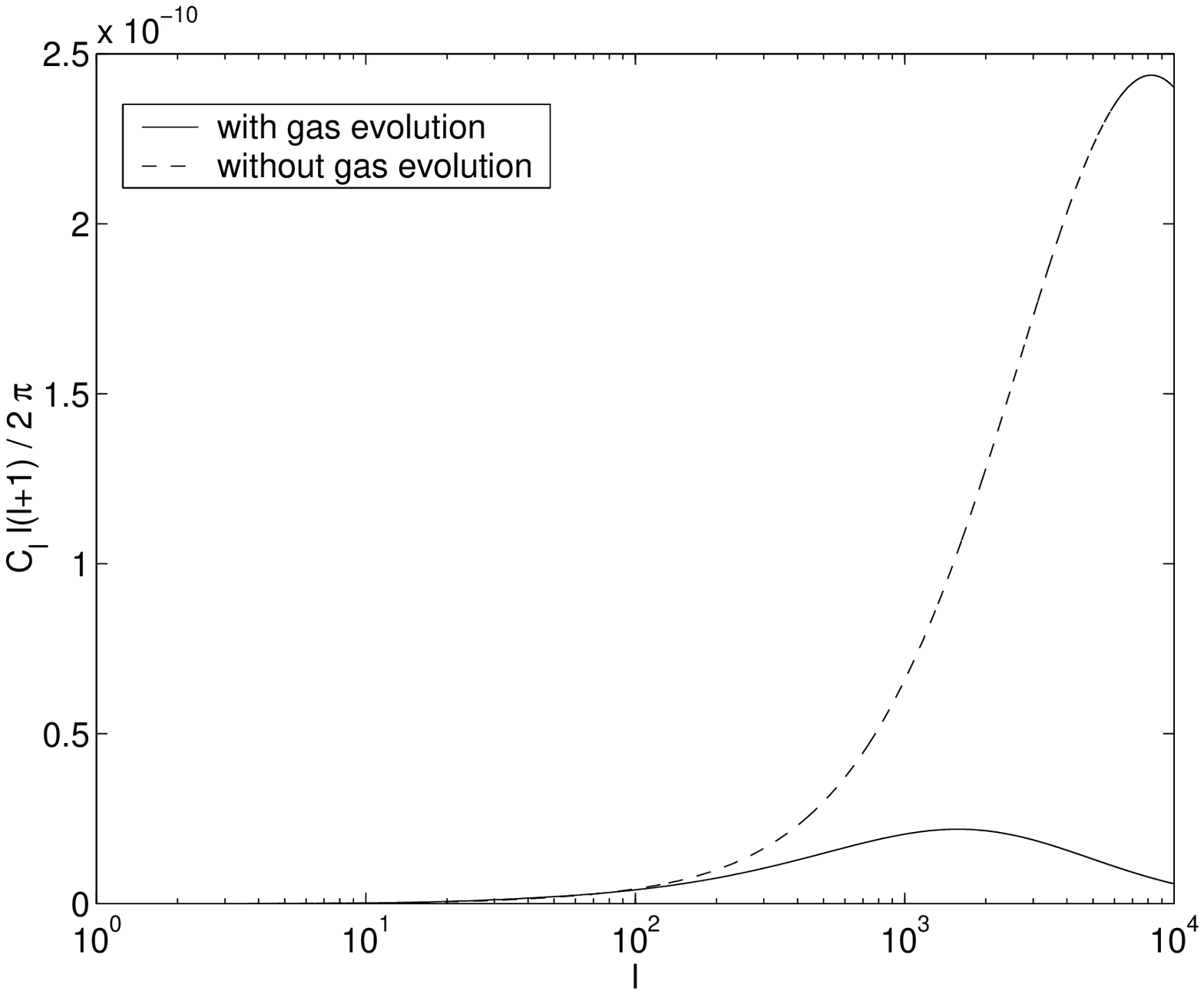, height=6.4cm, width=6.2cm, clip=}
\epsfig{file=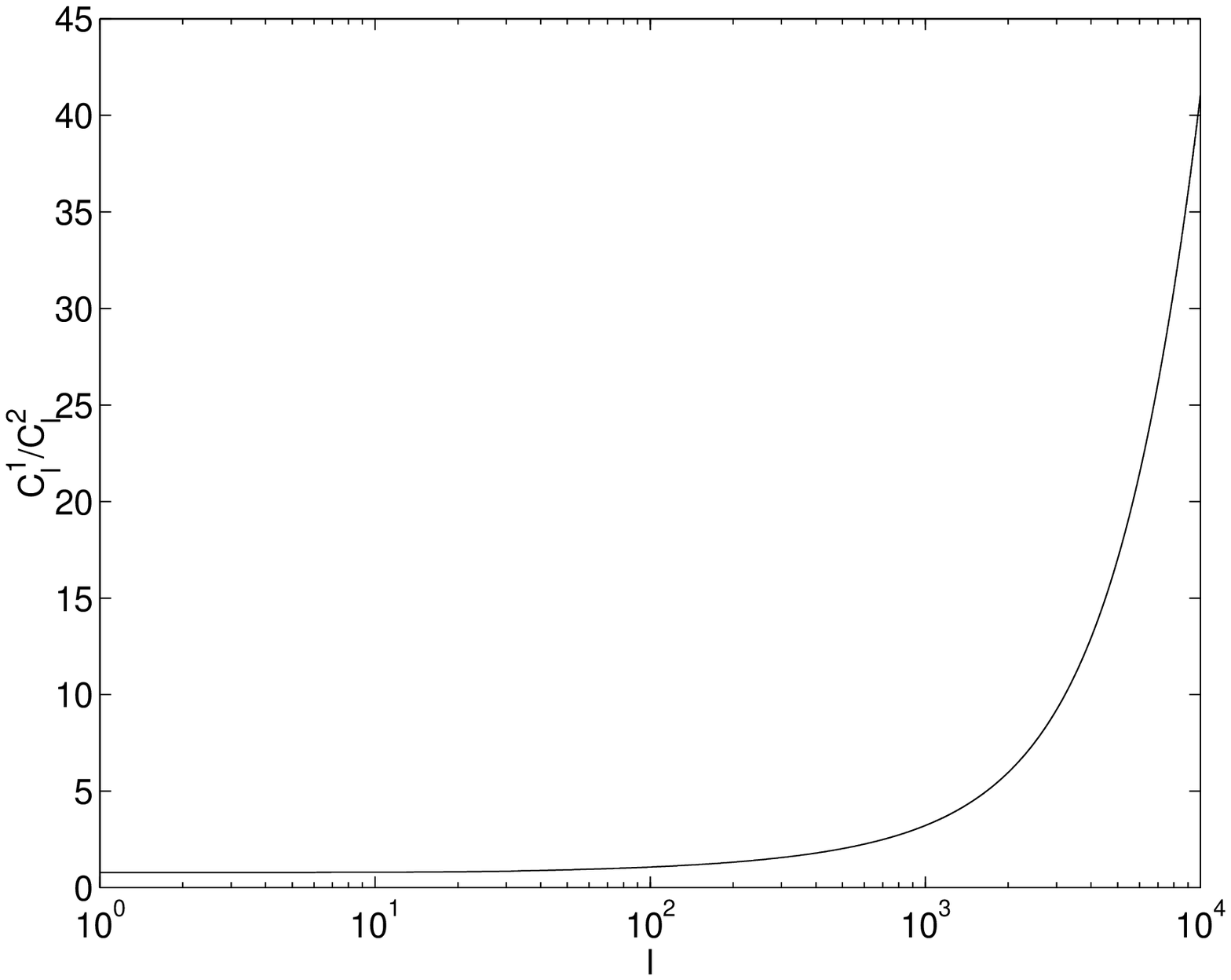, height=6.4cm, width=6.2cm, clip=}
\epsfig{file=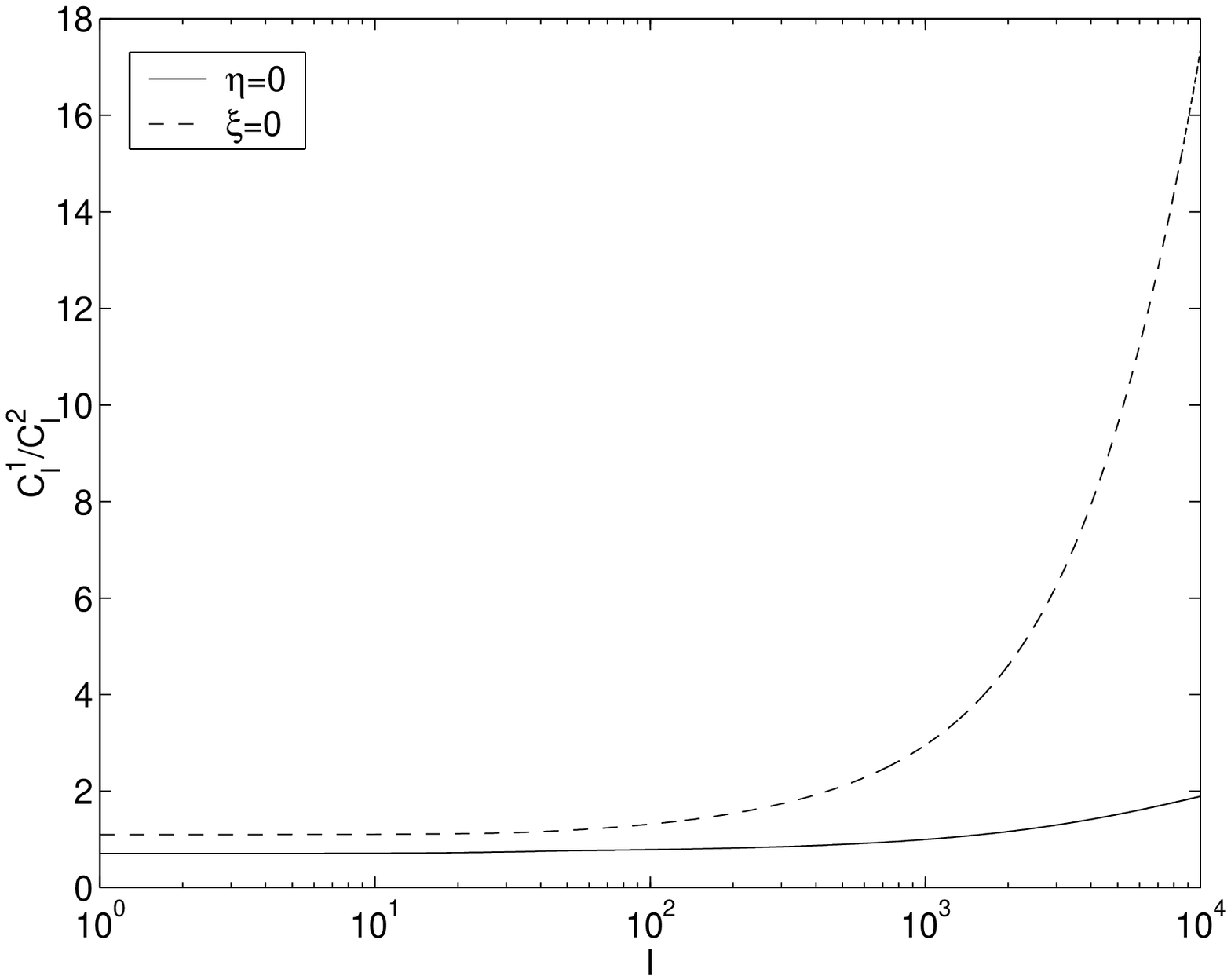, height=6.4cm, width=6.2cm, clip=}
\caption{The SZ angular power spectrum in the 
$\Lambda$CDM
model, with gas evolution (upper left panel) and compared to a non-evolving 
gas scenario (upper right panel). 
The lower left panel depicts the ratio 
between the two power spectra; the lower right panel describes the ratio 
between the power spectrum obtained in the non-evolving IC gas scenario and 
two 'partial' IC gas evolving models.}
\label{fig:gasevol}
\end{figure}

\newpage

\end{document}